\documentclass[11pt]{article}
\usepackage{moriond}
\usepackage{amsmath,amssymb}

\def\be{\begin{equation}}
\def\ee{\end{equation}}
\def\bea{\begin{eqnarray}}
\def\eea{\end{eqnarray}}

\newcommand{\Photo}{\medskip\includegraphics[height=35mm]{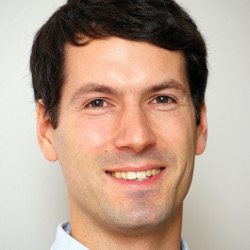}}

\begin{document}
\vspace*{4cm}
\title{CONSTRAINTS ON NEW PHYSICS FROM RARE (SEMI-)LEPTONIC {\em B}~DECAYS \footnote{%
Talk given at Rencontres de Moriond -- QCD and High Energy Interactions, 2013.
}}

\author{David M. Straub}

\address{PRISMA Cluster of Excellence \& Mainz Institute for Theoretical Physics,\\
Johannes Gutenberg University, 55099 Mainz, Germany}

\maketitle\abstracts{%
The measurements of rare leptonic and semi-leptonic $B$ decays performed at the LHC, not showing any significant deviations from the Standard Model expectations, put strong constraints on new physics. A brief account of the impact of recent measurements is given. As a concrete example, models with partial compositeness are discussed, where tree-level flavour-changing $Z$ couplings lead to potentially visibile effects in rare $B$ (and $K$) decays that will be scrutinized experimentally in the near future.
}




\section{Introduction}

The first LHC run has brought strong new constraints on physics beyond the Standard Model (SM). In the flavour sector, this is particularly true for new physics (NP) in $b\to s$ transitions, due to a first measurement of $B_s\to\mu^+\mu^-$ and improved measurements of $B\to K\mu^+\mu^-$ and $B\to K^*\mu^+\mu^-$ \cite{Aaij:2012vr,Aaij:2012nna,Aaij:2013iag}, including angular observables. While no significant deviations from the SM expectations have been found, the combined experimental and theoretical uncertainties still allow for NP effects of the order of 50\% of the SM contribution in many observables. It is therefore important not only to assess the impact of the improved constraints on NP models, but also to study in which observables visible deviations from the SM are still possible. After giving a brief, model-independent account of the impact on the most important classes of NP contributions in sec.~\ref{sec:mi}, a concrete model will be discussed in sec.~\ref{sec:z}, identifying the most promising observables to test it in the future.

\section{Model-independent considerations}\label{sec:mi}

In many concrete NP models (such as the MSSM at low $\tan\beta$ or models with partial compositeness), NP effects in the above-mentioned decays arise mainly through the operators\footnote{NP effects in the chromomagnetic operators $O_8^{(\prime)}$ also enter by mixing under renormalization with $O_7^{(\prime)}$.}
\begin{align}
\label{eq:O7}
O_7^{(\prime)} &= \frac{m_b}{e}
(\bar{s} \sigma_{\mu \nu} P_{R(L)} b) F^{\mu \nu},
&
O_{10}^{(\prime)} &=
(\bar{s} \gamma_{\mu} P_{L(R)} b)( \bar{\ell} \gamma^\mu \gamma_5 \ell)\,,
\end{align}
in the effective Hamiltonian
$$\mathcal H_\text{eff} = - \frac{4\,G_F}{\sqrt{2}} V_{tb}V_{ts}^* \frac{e^2}{16\pi^2}
\sum_i
(C_i O_i + C'_i O'_i) + \text{h.c.}$$
In these models, the contributions to the operators $O_9^{(\prime)}$ are typically small since $Z$-mediated effects are accidentally suppressed by the small vector coupling of the $Z$ to charged leptons.

In the case of the magnetic operators $O_7^{(\prime)}$, the impact of recent LHC measurements is mild, in view of existing constraints from the exclusive and inclusive radiative decays. The main new information is a reduced allowed size of $\text{Im}(C_7)$, stemming from the measurement of the branching ratio and $F_L$ in $B\to K^*\mu^+\mu^-$ at low $q^2$ \cite{Altmannshofer:2011gn,Altmannshofer:2012az}.

\begin{figure}
\centering\includegraphics[width=0.7\textwidth]{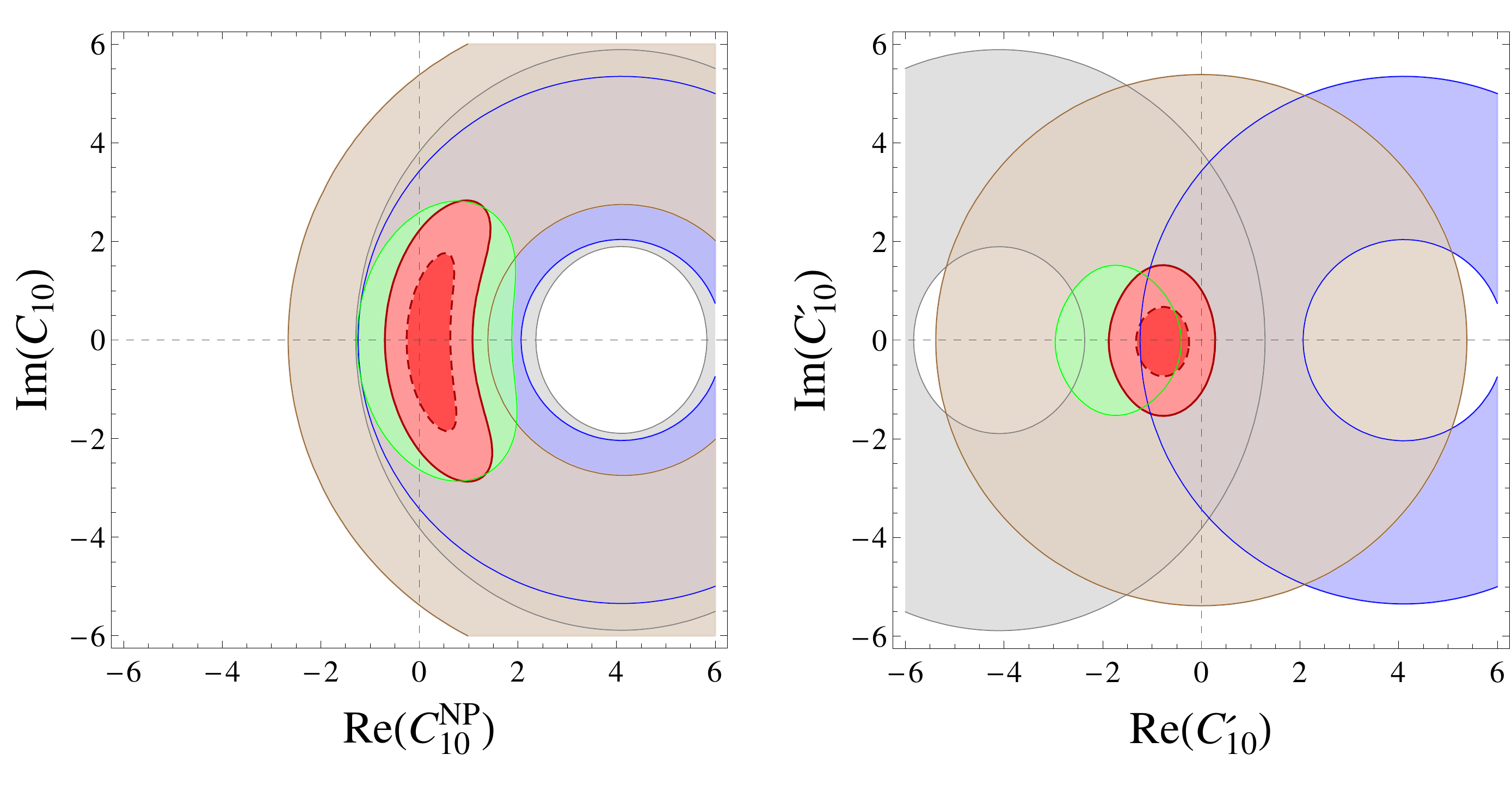}
\caption{Individual $2\sigma$ constraints on the NP contribution to $C_{10}$ and $C_{10}'$ from 
$B\to X_s\ell^+\ell^-$ (brown), $B\to K^*\mu^+\mu^-$ (green), $B\to K\mu^+\mu^-$ (blue) and $B_s\to\mu^+\mu^-$ (gray) as well as combined 1 and $2\sigma$ constraints (red).}
\label{fig:C10}
\end{figure}

The impact on allowed NP contributions to $O_{10}^{(\prime)}$ is much stronger. The combination of $B\to K^*\mu^+\mu^-$ and $B\to K\mu^+\mu^-$ measurements leads to a strong constraint on $C_{10}$ and $C_{10}'$ that is shown in fig.~\ref{fig:C10}. In the case of $C_{10}'$, the recent measurement of the angular observables $S_3$ and $A_9$ in $B\to K^*\mu^+\mu^-$, which were found to be in agreement with the SM, had a significant impact on CP-conserving and CP-violating contributions.\footnote{Note that the new $A_9$ measurement \cite{Aaij:2013iag} at low and high $q^2$ was included in the updated plot prepared for these proceedings.}
The operators $O_{10}^{(\prime)}$ can be generated by $Z$ penguin diagrams and are particularly large in models where flavour-changing $Z$ couplings are present at tree-level. This is the case in models with partial compositeness, as will be discussed in the next section.

In models with an extended Higgs sector, in particular the MSSM at large $\tan\beta$, also the (pseudo-)scalar operators can be relevant,
\begin{align}
O_S^{(\prime)} &= 
\frac{m_b}{m_{B_s}} (\bar{s} P_{R(L)} b)(  \bar{\ell} \ell)\,,
&
O_P^{(\prime)} &=
\frac{m_b}{m_{B_s}} (\bar{s} P_{R(L)} b)(  \bar{\ell} \gamma_5 \ell)\,.
\end{align}
While the contributions of $O_{S,P}^{(\prime)}$ to $B\to K^*\mu^+\mu^-$ turn out to be numerically irrelevant \cite{Altmannshofer:2008dz}, they can lead to large enhancements in $B_s\to\mu^+\mu^-$. Indeed, the LHCb measurement leads to a strong bound on the MSSM with large $\tan\beta$ and a light pseudoscalar Higgs boson even for Minimal Flavour Violation (see e.g.\ \cite{Altmannshofer:2012ks}).

\section{{\em Z} penguins and partial compositeness}\label{sec:z}

Apart from supersymmetry, models with a composite Higgs boson (and their dual extra-di\-men\-sion\-al theories) arguably constitute the most well-motivated solution to the gauge hierarchy problem. In these models, to avoid excessive flavour-changing neutral currents (FCNCs), one generates fermion masses by the mechanism of partial compositeness. One consequence is that the couplings of quarks to the $Z$ boson generically receive non-universal corrections that
are constrained by electroweak precision tests (EWPT) and lead to tree-level FCNCs.
To avoid the EWPT constraints, discrete symmetries can be imposed that in turn lead to characteristic patterns of effects in $Z$-mediated FCNCs.

\begin{figure}
\centering
\includegraphics[width=0.3 \textwidth]{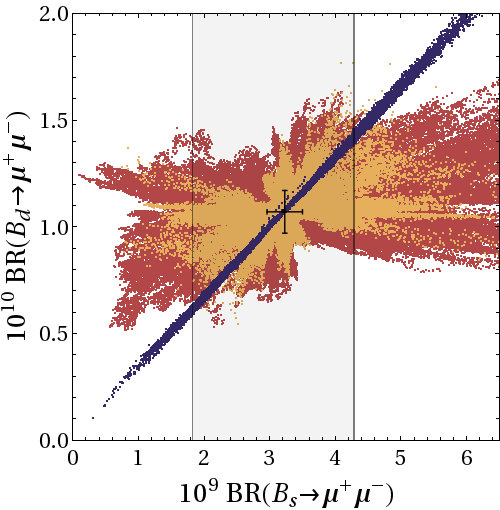}%
\hspace{0.03\textwidth}%
\includegraphics[width=0.3 \textwidth]{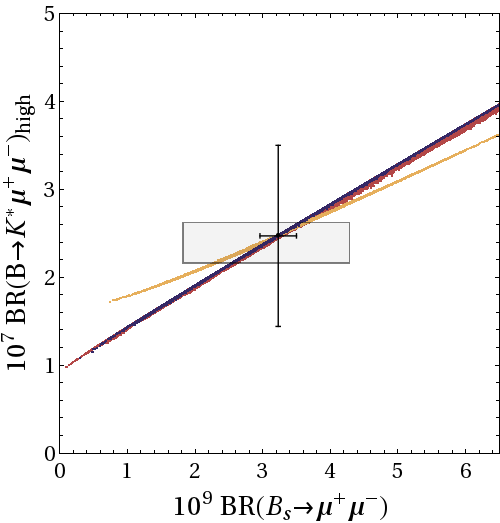}
\hspace{0.03\textwidth}%
\includegraphics[width=0.3 \textwidth]{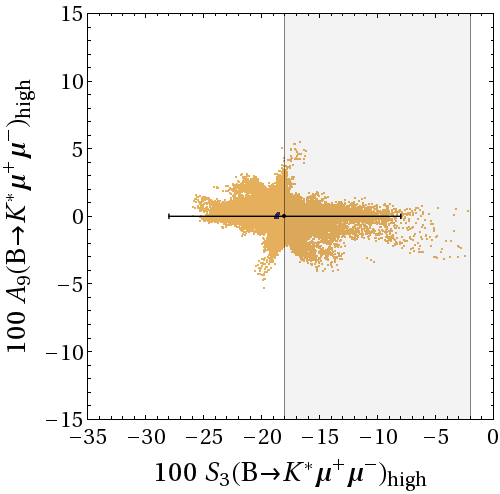}
\caption{Correlations between rare $B$ decay observables in three models with partial compositeness: the anarchic triplet model (yellow), the anarchic bidoublet model (red) and the bidoublet model with $U(2)^3$ left-handed compositeness (blue). For details see ref.~8.}
\label{fig:BR_all}
\end{figure}

Recently \cite{Straub:2013zca}, a complete numerical study of $Z$-mediated FCNCs in models with partial compositeness was performed for two different choices of composite fermion representations and for two choices of the flavour structure of the composite sector: anarchy and a minimally broken $U(2)^3$ flavour symmetry \cite{Barbieri:2012uh,Barbieri:2012tu}.

It was found that  in all the scenarios, several branching ratios of rare $K$ and $B$ decays can exhibit visible effects at current and planned experiments. Clear-cut correlations between various observables can be used to distinguish the different models on the basis of the pattern of observed deviations from the SM expectations.

An example is shown in fig.~\ref{fig:BR_all}: the NP contributions to $B_s\to\mu^+\mu^-$ and $B\to K^*\mu^+\mu^-$ through the operators $O_{10}^{(\prime)}$ lead to effects that are comparable to the current experimental precision. The correlation in the left-hand plot present in the $U(2)^3$ model is due to the flavour symmetry relating $b\to d$ and $b\to s$ transitions. The correlation in the centre plot is due to the fact that only $C_{10}$ {\em or} $C_{10'}$ (corresponding to only left-handed or right-handed flavour-changing $Z$ couplings) contribute in all models. In the right-hand plot, one can see that in a model with right-handed flavour-changing $Z$ couplings, also non-zero contributions to the angular observables $S_3$ and $A_9$ can be generated, although in this concrete model they turn out to be small, such that the recent improved measurement by LHCb (not shown in the plot) does not have an impact yet.

\section{Conclusions}

The failure of NP to show up in rare $B$ decays sensitive to the $b\to s$ transition has led to strong constraints on physics beyond the SM. In a large class of NP models, the most significant such constraints are
\begin{itemize}
\item Less room for a CP-violating contribution to the magnetic operator $O_7$.
\item Less room for NP in scalar and pseudoscalar operators $O_{S,P}^{(\prime)}$. This is particularly relevant for the MSSM, where  it effectively sets an upper bound on the combination $\tan^3\beta/M_A^{2}$.
\item Less room for NP in the semi-leptonic operators $O_{10}^{(\prime)}$, which can be generated for instance by $Z$ penguin diagrams.
\end{itemize}
In the latter case, the impact of the recent first LHCb measurement of the angular CP asymmetry $A_9$ was shown in the right-hand plot of fig.~\ref{fig:C10}.

In sec.~\ref{sec:z}, a concrete NP framework was discussed where tree-level contributions to the operators $O_{10}$ or $O_{10}'$ are generated. Future improved measurements, complemented with progress on the theoretical uncertainties, will allow to probe a significant part of the parameter space of these models.

\section*{Acknowledgments}

I warmly thank the organizers for the invation to this conference. Thanks also to Wolfgang Altmannshofer and Christoph Bobeth for useful discussions.
This work was supported by the Advanced Grant EFT4LHC of the European Research Council (ERC), and the Cluster of Excellence {\em Precision Physics, Fundamental Interactions and Structure of Matter\/} (PRISMA -- EXC 1098).

\end{document}